\documentclass[10pt]{article}

\usepackage{amsmath}
\usepackage{amssymb}

\usepackage{graphicx}

\usepackage{mdframed}

\usepackage{cite}

\usepackage{color} 

\usepackage{multirow}

\usepackage[labelfont=bf,labelsep=period,justification=raggedright]{caption}

\makeatletter
\renewcommand{\@biblabel}[1]{\quad#1.}
\makeatother

\date{}

\begin{document}

\begin{flushleft}
{\Large
\textbf{Geometry shapes evolution of early multicellularity}
}
\\
Eric Libby$^{1 \ast}$, 
William Ratcliff $^{2}$,
Michael Travisano $^{3}$,
Ben Kerr $^{4}$
\\
\bf{1} Santa Fe Institute, Santa Fe, New Mexico, United States 
\\
\bf{2} School of Biology, Georgia Institute of Technology, Atlanta, Georgia, United States
\\
\bf{3} Department of Ecology, Evolution, and Behavior, University of Minnesota, St. Paul, Minnesota, United States
\\
\bf{4} Department of Biology and BEACON Center, University of Washington, Seattle, Washington, United States
$\ast$ E-mail: elibby@santafe.edu
\end{flushleft}

\section*{Abstract} 
Organisms have increased in complexity through a series of major evolutionary transitions, in which formerly autonomous entities become parts of a novel higher-level entity.  One intriguing feature of the higher-level entity after some major transitions is a division of reproductive labor among its lower-level units. Although it can have clear benefits once established, it is unknown how such reproductive division of labor originates. We consider a recent evolution experiment on the yeast {\it Saccharomyces cerevisiae} as a unique platform to address the issue of reproductive differentiation during an evolutionary transition in individuality. In the experiment, independent yeast lineages evolved a multicellular ``snowflake-like'' cluster form in response to gravity selection. Shortly after the evolution of clusters, the yeast evolved higher rates of cell death. While cell death enables clusters to split apart and form new groups, it also reduces their performance in the face of gravity selection. To understand the selective value of increased cell death, we create a mathematical model of the cellular arrangement within snowflake yeast clusters. The model reveals that the mechanism of cell death and the geometry of the snowflake interact in complex, evolutionarily important ways. We find that the organization of snowflake yeast imposes powerful limitations on the available space for new cell growth. By dying more frequently, cells in clusters avoid encountering space limitations, and, paradoxically, reach higher numbers. In addition, selection for particular group sizes can explain the increased rate of apoptosis both in terms of total cell number and total numbers of collectives. Thus, by considering the geometry of a primitive multicellular organism we can gain insight into the initial emergence of reproductive division of labor during an evolutionary transition in individuality.

\section*{Introduction}
Organisms have increased in complexity through a series of major evolutionary transitions, in which formerly autonomous entities become parts of a novel higher-level entity \cite{Smith1997,Calcott2011,Okasha2005,Bonner1998,Michod1997}. Examples of this transition include the evolution of multicellular organisms from unicellular ancestors and eusocial ``superorganisms" from multicellular ancestors.  One of the primary benefits ascribed to major evolutionary transitions is the potential for the higher-level entity to evolve division of labor among its lower-level units \cite{Smith1997,Calcott2011}. A salient form of this is reproductive division of labor as found in germ-soma differentiation in multicellular organisms and worker/queen roles in eusocial insects \cite{Michod2005a,Koufopanou1994,Bendich2010,Sherman1995}. While reproductive specialization is not strictly required for division of labor to provide a fitness benefit to the higher-level entity, it has evolved repeatedly in independent lineages \cite{Grosberg2007,Herron2013,Jarvis1981,Crespi1992,Duffy1996}. Upon a superficial glance, the existence of such reproductive self-sacrifice seems to present an evolutionary paradox. How would such a self-destructive tendency be favored by a process (natural selection) that places a high premium on survival and reproduction? The resolution of the paradox generally involves a situation in which the self-sacrifice improves the fitness of the higher-level unit \cite{Solari2006,Engelberg2006, Bendich2010,Koufopanou1994,Jacobson1997,Michod2005a,Michod2005b,Okasha2009}.  For ease of discussion, let us call the lower-level entities ``particles" and the higher-level entities ``collectives."  Suppose the altruistic action of some particles allows other particles in their collective to found new collectives at a higher rate. If these founding particles possess a tendency for self-sacrifice (which can occur if particles have high relatedness within collectives), then reproductive division of labor within collectives can evolve. We emphasize that such altruism must occur in a strict subset of particles within the collective and requires the plastic or stochastic expression of phenotypic traits at the particle level. 
Thus, while the logic for the evolution of reproductive self-sacrifice is sound, the mechanistic underpinnings could be complex. The precise way in which such differentiation evolves and its presence in the early stages of major transitions are largely unknown.
\par
A recent evolution experiment on the yeast {\it Saccharomyces cerevisiae} has provided a unique platform to address the issue of reproductive differentiation during an evolutionary transition in individuality \cite{Ratcliff2012}. In this experiment, populations of unicellular yeast were periodically exposed to a selective regime that rewarded cells that sank quickly in test tubes. During this setting, cells in clusters sink more quickly than independent cells, incentivizing group formation. Cluster-forming phenotypes evolved repeatedly via the retention of cell-cell connections after mitotic reproduction. These group-forming types outcompeted their unicellular ancestors, driving them to extinction in all 10 replicate populations within 60 days \cite{Ratcliff2012}. Clusters grew in size until the resulting physical strain caused them to fragment, yielding a form of group reproduction. Thus, the yeast evolved group formation and reproduction {\it de novo}. Interestingly, these yeast clusters soon evolved a secondary trait: a higher rate of cellular programmed death (hereafter referred to as apoptosis). Why would a higher rate of cellular suicide, an ostensibly costly trait, be favored by natural selection? 
\par
A higher rate of apoptosis might have evolved because it increases collective-level reproduction. Since each cell in the group is connected solely to its parent and offspring cells \cite{Ratcliff2012}, it only takes a single break in any connection to produce two distinct collectives. Both physical strain and cell death can create such breaks and, consequently, increase the number of groups. Selection for a greater number of clusters could therefore lead to the evolution of higher rates of apoptosis. Yet, the problem is that the selective regime seemingly rewards large clusters, not large {\it numbers} of clusters.  The apoptotic mechanism of group reproduction is simultaneously a mechanism of group-size reduction. While there may be a benefit for groups to reproduce (to reduce the risk of not being transferred due to random sampling error), as groups divide they become smaller and sink less quickly, making them less competitive against larger groups.  It would appear that an optimal strategy would be for groups to grow as large as possible and divide infrequently. In contrast, when selection for large groups is stronger (requiring faster settling), groups evolve higher rates of apoptosis and produce proportionally smaller propagules \cite{Ratcliff2012}.
\par
To address this conundrum, we create a mathematical model of the cellular arrangement within snowflake yeast clusters. The model reveals how the geometric structure of the cells in the cluster interacts with apoptosis to affect the number and size of group offspring.  We find that the organization of snowflake yeast imposes powerful limitations on the available space for new cell growth. By dying more frequently, cells in clusters avoid encountering space limitations, and, paradoxically, reach higher numbers. Finally, we demonstrate that selection for particular group sizes can explain the increased rate of apoptosis both in terms of total cell number and total numbers of collectives. Thus, this model provides an explanation for the evolution of reproductive self-sacrifice and the emergence of reproductive division of labor during an evolutionary transition in individuality.


\section*{Model \& Results}
\subsection*{Group growth and reproduction}
\begin{figure}[!ht]
\begin{center}
\includegraphics[width=.9 \textwidth]{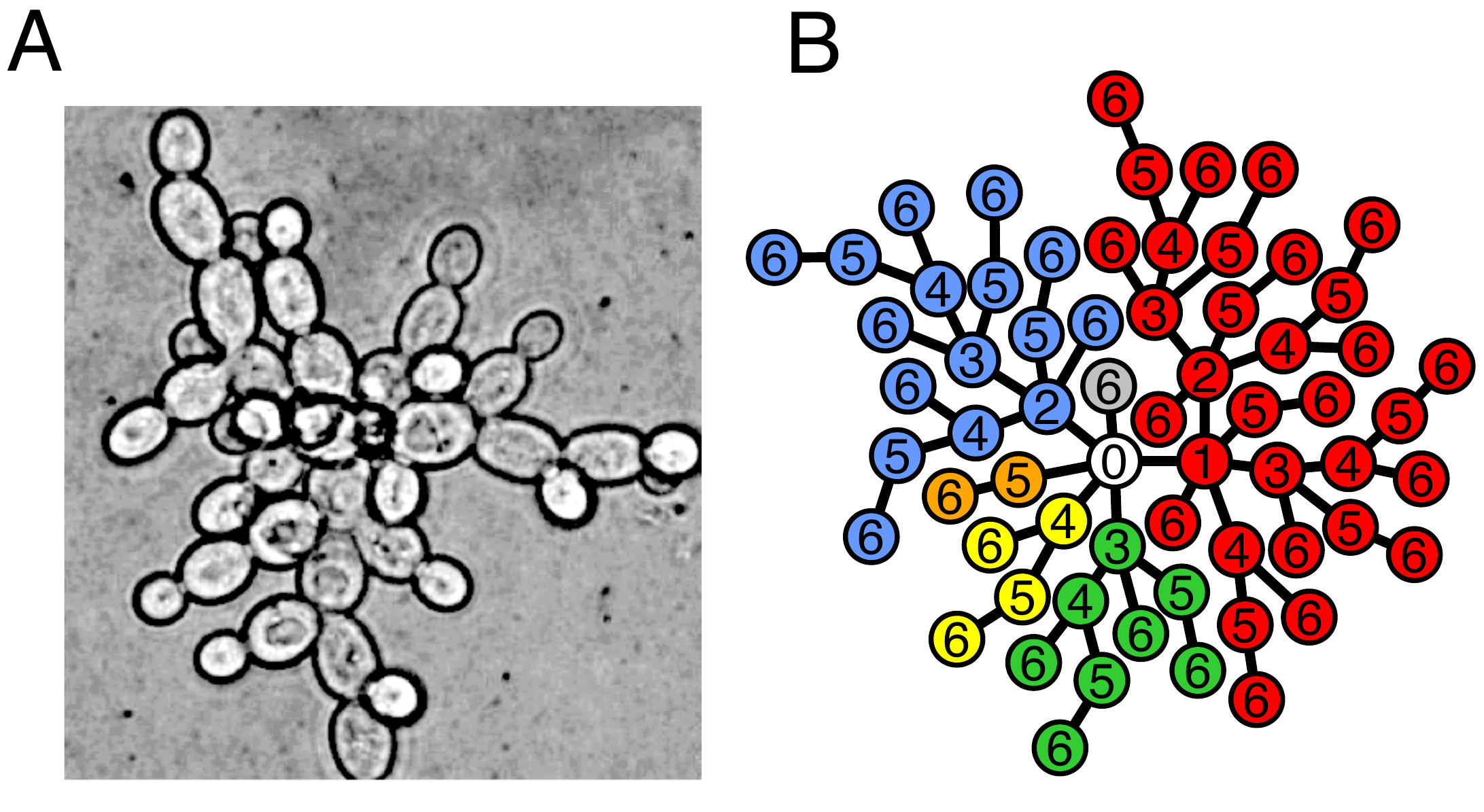}
\caption{{\bf Groups as trees}. {\bf A)} Photograph of a cross-section of the yeast snowflake phenotype shows the branching morphology. {\bf B)} Simulated group growth from a single cell ({\it Node 0}) after 6 rounds of cell divisions. The different colors represent different branches emanating from {\it Node 0}.  The numbers inside nodes represent the cell division of their birth.
}
\label{Tree}
\end{center}
\end{figure}

We describe the structure of evolved yeast groups by a tree graph in which nodes represent cells and edges represent physical attachments between cells (Figure \ref{Tree}). When a cell divides, its corresponding node in the tree gains an edge to a newly created node. For simplicity we begin the tree with only one node which represents the first mutant yeast cell to have the capacity to form groups, call it {\it Node 0}. Each time {\it Node 0} divides it generates a branch which will continue to grow independently. Initially, we assume that all cells reproduce and do so at the same rate. Thus, with each successive division the tree doubles its nodes, i.e. the group doubles in the number of cells. After $n$ divisions the branches from {\it Node 0} will be composed of $2^{n-1}, 2^{n-2},\ldots 2^0$ cells depending on which cell division the branch was initiated (and the total number of cells in the tree is $ 1+ \sum_{i=0}^{n-1} {2^i} = 2^n$). 
\par
A consequence of this spatial structure is that if a link/edge between two cells/nodes is severed then it will result in two distinct groups, i.e. the group reproduces. Since both physical strain and cell death lead to group reproduction, we can view these as mechanisms for severing an edge between two nodes. The location of the severed edge plays a significant role in determining the sizes of the resulting group offspring. If an edge in the periphery is severed then one of the resulting groups will be composed of only a single cell. In contrast, severing more central edges will result in less asymmetry between offspring groups. 
\par
The manner by which cells die determines whether a severed edge is more likely to be in the periphery or the center. If cell death is completely random such that the centermost cells are just as likely to die as newly created cells then the severed edge is more likely to be peripheral under the scenario in which the tree doubles every division. This is because at any time $50\%$ of the tree is newly created. As a result, there is a $50\%$ chance that death of a random cell will yield a ``group" that is one (dead) cell in size by breaking its one and only link to the tree. The expected sizes of the offspring groups after $n$ divisions are $\frac{n}{2}$ and $2^n-\frac{n}{2}$, and the ratio of the smaller offspring to the parent is less than $0.5\%$ after 10 rounds of cell divisions (the ratio, $P(n)$, after $n$ division is $\frac{n}{2^{n+1}}$). Such a small group may not be able to grow large enough to survive the selective regime and could be excluded from future growth and reproduction. 
\par
If cell death is not completely random but rather related to age then central edges would be more likely to be severed. In the case that the oldest cell ({\it Node 0}) dies, the sizes of the resulting offspring groups will depend on which link is severed. Each link of {\it Node 0} corresponds to one of its branches with $2^{n-1}, 2^{n-2},\ldots,2^0$ cells. Without a bias as to which link is severed the smaller offspring would be expected to have $\frac{2^n-1}{n}$ cells. The ratio of this offspring to the parent after $n$ divisions, $P(n)$, is $\frac{1}{n}(1-\frac{1}{2^n})$. After $10$ cell divisions $P(n)$ is approximately $10\%$ which is 20 times larger than when cell death is completely random. Thus, weighting death towards older cells increases the size of the smaller offspring.
\par
Experimental observations of early group offspring in the yeast system suggest that the smaller offspring may be closer to $20-40\%$ of the size of the parent \cite {Ratcliff2012}. To see how link severance via cell death can achieve such values, we consider again the death of {\it Node 0} which yielded less offspring asymmetry than random cell death. The oldest branch of {\it Node 0}, created at the first division, is the only branch greater than 40\% of the tree size-- it is half of the size of the whole tree. The next oldest branch, created at the second division of  {\it Node 0}, is a fourth of the whole tree size. Each successive branch is half the size of the previous. If there is no bias in which branch becomes the offspring then the odds favor the $n-1$ branches that are much less than 40\%. Instead of unbiased link severing, it could be that links are severed according to the size of the branch they are supporting. Bigger branches may produce more strain on their links compared to smaller branches and, therefore, may break more easily. If the probability a link is severed is proportional to the size of the branch then the ratio of the smaller offspring to the parent after $n$ divisions, $P(n)$, is $\frac{1}{3}( 1+ \frac{1}{ 2^n})$ which approaches $\frac{1}{3}$ as $n$ increases. This matches experimental observations more closely and suggests that reproduction via cell death may be biased both in which cells die and which links are severed. 
\par
\subsection*{Optimal strategies}
To get a rough understanding of how the experimental regime selects for different rates and mechanisms of cell death, we briefly abandon considerations of geometry in this section and ask how a cluster might optimally divide. When a cluster splits as a result of cell death, it yields both new and smaller clusters. Thus, division simultaneously affects cluster reproduction and the prospects for viability under settling selection.  How should a cluster balance fecundity against survival?  Because the experimental regime of Ratcliff {\it et al.} \cite{Ratcliff2012} includes a growth phase followed by a selection event, we seek a framework in which division can depend on time within the growth phase. Because cluster growth and division changes the size of clusters, we also seek a framework in which the splitting strategy can be size-dependent. Here we use a dynamic programming approach \cite{Mangel1988, Houston1999} to explore optimal division strategies.
\par
We denote the probability a cluster of $x$ cells survives settling selection as $S(x)$. Since larger groups settle faster than smaller ones, we assume $S(x)$ is a non-decreasing function. In addition, we assume that division and growth of clusters occur for $T$ time steps prior to settling selection. We define $F(x,t)$ to be the maximal reproductive output for a cluster of size $x$ ($x \geq 0$) at time $t$ ($0 \leq t \leq T$). Thus, if fitness is measured in terms of number of clusters $F(x,T)=S(x)$, and if fitness is measured in terms of the number of cells $F(x,T)=x S(x)$. Over each time step (from $t$ to $t+1$, where $t \in \{0,1,2,... T-1\}$), we assume that clusters divide and then grow. Specifically, a cluster of $x$ cells at time point $t$ splits into two clusters of sizes $p$ and $x-p$ (where $0 \leq p \leq {x/2}$). We note that this includes the case where the cluster does not divide (i.e., $p=0$). After division, the new clusters grow according to the function $G(x)$; that is, a cluster that starts with $x$ cells ends the time step with $G(x)$ cells. For instance, if every cell in a cluster doubles over a time step, then $G(x)=2x$. We have the following backwards recursion for maximal reproductive output:
\par
\begin{equation}
\label{dprecursion}
F(x,t)=\max_{0 \leq p \leq {x/2}}(F(G(p),t+1)+F(G(x-p),t+1))
\end{equation}
\par

Suppose fitness is measured in terms of the number of clusters that survive selection (i.e., $F(x,T)=S(x)$). In the Supplement, we prove that if $G(x)=ax$, where $a$ is some positive integer greater than 1 and $S(x)$ is concave down (e.g., $S(x)=1-e^{-c x}$), then the optimal strategy is always to divide into halves (or as close to halves as possible). This result would predict the evolution of cell death mechanisms that produce equal sized group offspring. Generally in this case, higher splitting rates would be favored and cell death may be one way to accomplish this. In addition, it would also predict that groups remain small in size as there is always an advantage to splitting in half. However, upon reflection, there are several assumptions that could be challenged.  If $S(x)$ is convex over some range of $x$, then it can be optimal not to divide at all (at least for some sizes; see Supplement). If fitness is measured in terms of the number of cells (i.e., $F(x,T)=xS(x)$) rather than the number of clusters, it can be optimal not to divide even when $S(x)$ is strictly concave (see Supplement).  Finally, the assumption $G(x)=ax$ is a clear oversimplification, as this implies that a cluster can grow exponentially without any restrictions imposed by its size.  In the next section, we consider how basic geometry constrains cluster growth and then in the following section we return to the question of the adaptive value of increasing cell death rate.

\subsection*{Growth constraints}
\begin{figure}
\begin{minipage}[c][1\width]{0.5\textwidth}%
\includegraphics[clip,width=1\textwidth]{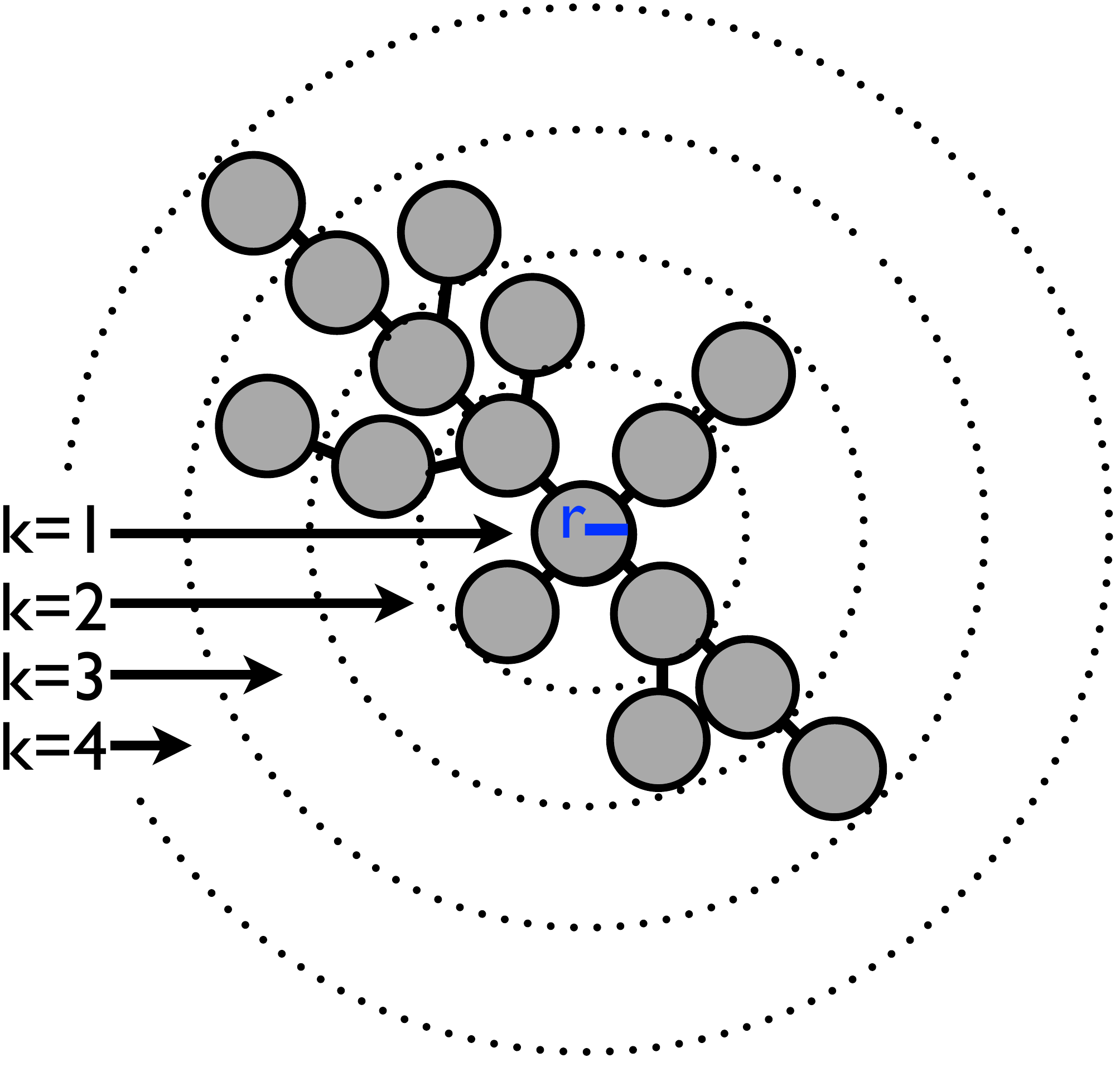}
\end{minipage}%
\begin{minipage}[c][1\width]{0.5\textwidth}%
\begin{center}
\begin{tabular}{|c|c|c|c|c|c|c|c|}
\hline
\multirow{2}{*}{Divisions ($n$)} & \multicolumn{6}{|c|}{Shells ($k$)} \\
  & 0 & 1& 2 & 3 & 4 & 5 \\
\hline
 0 & 1 & 0& 0 & 0 & 0 & 0  \\
1 & 1 & 1& 0 & 0 & 0 & 0  \\
2 & 1 & 2& 1 & 0 & 0 & 0  \\
3 & 1 & 3& 3 & 1 & 0 & 0  \\
4 & 1 & 4& 6 & 4 & 1 & 0  \\
5 & 1 & 5& 10 &10 & 5 & 1\\
\hline
\end{tabular}
\end{center}%
\end{minipage}
\caption{{\bf Volume constraints to tree growth} {\bf (Left)} A model of the growing tree with {\it Node 0} at the center and shells of nodes surrounding it. Each cell is a sphere with radius r and the edges are only shown to make relationships clear-- edge length is effectively 0. {\bf (Right)} After $n$ divisions, each shell $k$ contains ${n \choose k}$ cells exclusively. }
\label{Vol}
\end{figure}

Until now, we have operated under the unrealistic assumption that there are no limits to cell division. While each cell division increases the size and span of the group, it also fills the limited volume at the center. As this space gets crowded, cells lose both access to nutrients and room for further division. To determine how the tree geometry experiences volume constraints, we use a 3-dimensional model of growth in which cells occupy concentric orbitals/shells surrounding the central node, {\it Node 0} (Figure \ref{Vol}). By stretching the group along its longest diameter, this model maximizes the available space and sets an upper bound to the size capacity of the group. 
\par
We assume that each cell is an identical sphere with radius $r$. Cells occupy shells depending on how many links separate them from {\it Node 0}. For example, the third shell is filled with cells that are 3 links from the center. The offspring of a cell occupies the next shell and, conversely, its parent is in the previous shell. Each shell $k$ encloses a volume equivalent to a sphere with radius $R=r+2kr$. This volume ($\frac{4}{3}\pi R^3$) can hold at most $(1+2k)^3$ cells-- ignoring issues concerning the maximum packing of spheres. In the growing group, the actual number of cells within the volume of a shell is simply the total number of cells in each interior shell. For a given shell $k$ after $n$ divisions the total number of cells is $ \sum_{j=1}^k {n \choose j}$ (see Figure 2). Thus, the volume enclosed by shell $k$ is exceeded when the number of cell divisions $n$ satisfies: 
 \begin{equation}
 \label{VolEq}
\sum\limits_{j=1}^k {n \choose j}  > (1+2k)^3  
 \end{equation}
We calculate the lowest $n$ for which the volume bounded by each shell is exceeded and find shells 4-6 are the first to overflow at the twelfth division ($n=12$). Even if a cell in shell 4 could relocate to shell 3 there is no room available because the volume defined by shell 4 has been exceeded. While there is still space in the volume contained by shells 7-12, cells from the overcrowded volume cannot move here because they must remain connected to their parents in more interior shells.
\par
\begin{figure}
\includegraphics[width=.88\textwidth]{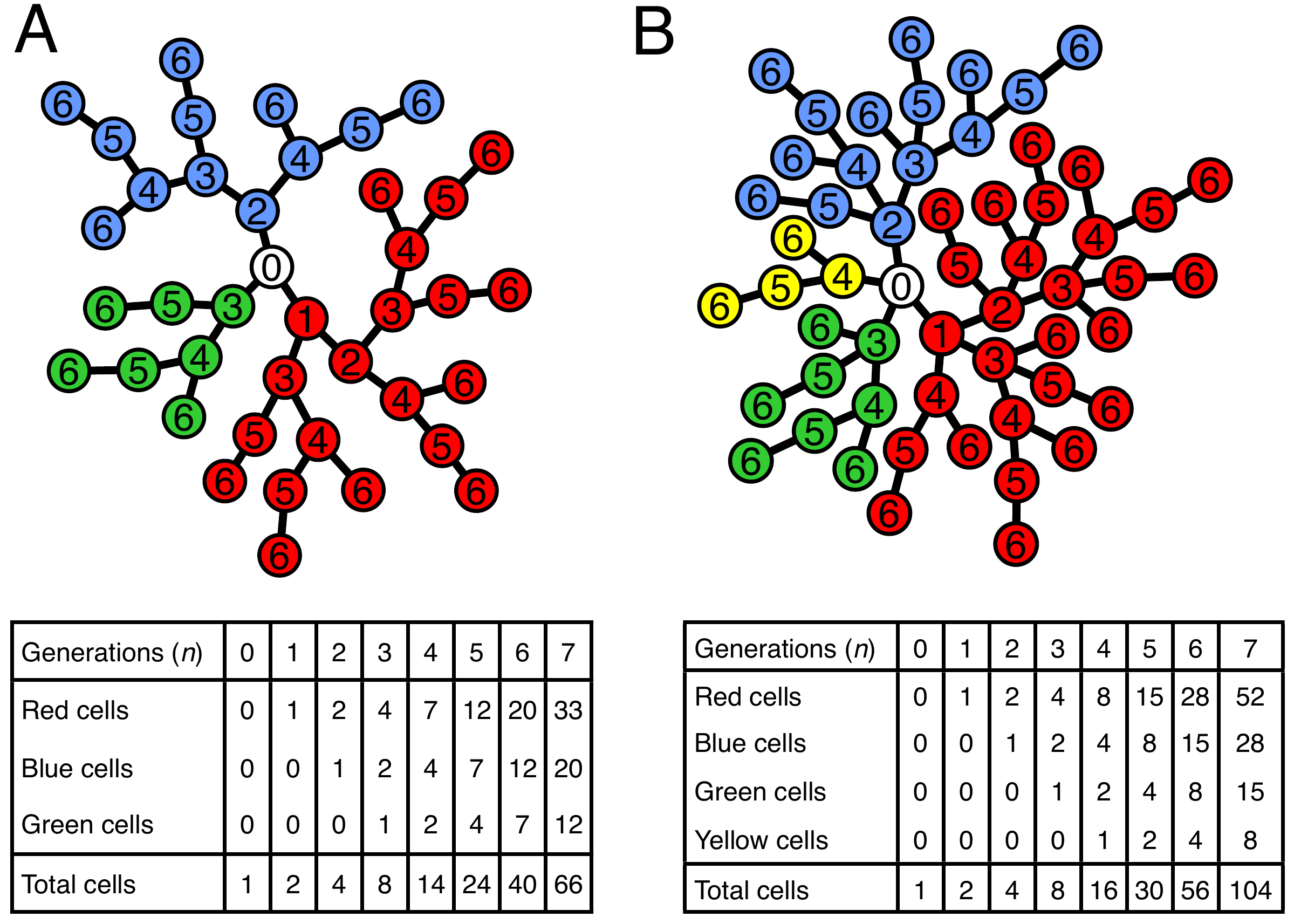}
\caption{{\bf Degree capped tree growth} {\bf A)} A model of a growing tree with {\it Node 0} at the center and a degree cap of 3. The numbers inside each node represent the generation of their birth while the colors denote the 3 different branches emanating from {\it Node 0}. The table below shows the total number of cells in each branch and the tree as a whole as a function of the number of generations. The number of nodes of each branch follow the same series: $1,2,4,7,\ldots$ described by the recursion $a_n=a_{n-1}+a_{n-2}+1$. This can be solved analytically to get $a_n = (\frac{5-3 \sqrt{5}}{10}) \big( \frac{1-\sqrt{5}}{2}  \big)^n+ (\frac{5+3 \sqrt{5}}{10}) \big( \frac{1+\sqrt{5}}{2} \big)^n -1$. {\bf B)} A model of a growing tree with a degree cap of 4. Similar to A) there is a recursive relationship for the number of nodes in a branch but it delves one more generation into the past, i.e. $a_n=a_{n-1}+a_{n-2}+a_{n-3}+1$. For both trees the total number of nodes in the tree is twice the number in the red branch.}
\label{Prune}
\end{figure}

In addition to the volume constraint, there may be constraints regarding how many attachments (edges) a single cell can have. Experimental observations of group structure find that most cells are attached to only a few cells ($<5$). If there is a limit to the number of attachments per node then this will alter the organization of a group (Figure \ref{Prune}). For example, a tree with maximum node degree of 3 will have just 3 branches emanating from {\it Node 0}. Instead of doubling with each generation, the number of nodes in a branch follows a recursion: $a_n = a_{n-1} + a_{n-2} + 1$, where $a_n$ is the number of nodes $n$ generations after the creation of the branch. 
Geometries with higher maximum node degrees (hereafter called ``degree capped") also feature recursive relationships such that in general, for a tree with degree cap of $m$, $a_n = a_{n-1} + a_{n-2} + \ldots + a_{n-m+1} + 1$ with the first $m$ values following $a_n=2^{n-1}$. This stems from an important distinction in trees with a degree cap: their size only increases with those cells who were created within the last $m$ generations. These recursive relationships relate cluster sizes with Fibonacci numbers such that trees without degree caps are simply Fibonacci sequences of infinite order. In all cases, the total number of nodes in the tree is simply twice the number in the largest branch.   
\par
As the distribution of cells in branches is altered by limiting the number of node attachments so, too, is the expected size of offspring groups. If {\it Node 0} dies and there is no bias to which link is severed then the expected offspring size as a proportion of the parent is $P(n)=\frac{1}{m}(1-\frac{1}{2 a_n})$, where $m$ is the degree limit and $a_n$ is the number in the largest branch. This quickly approaches $\frac{1}{m}$ which is much greater than the value $\frac{1}{n}$ found in trees without limits to the number of node attachments. Consequently, group offspring are more equal in size. In trees without degree caps, biasing which link is severed according to branch size increased the symmetry of group offspring such that the expected size of the smaller offspring is $33\%$ of the parent's size. By comparison, biasing link severance in trees with degree caps has less of an effect. The expected size of the smaller offspring is $38.2\%$ of the parent's size for a degree cap of 3 and $35.2\%$ for a degree cap of 4. While the most symmetric group reproduction is in trees with a degree cap of 3 and biased link severance, all trees with biased link severance produce an offspring that is between $33-38.2\%$ of the parent's size.  It should be noted that a degree cap of 2 can do better but it can only form filaments rather than snowflake-shaped clusters. 
\par
Not only do trees with degree caps produce more symmetric offspring but they experience less severe volume constraints. Since limiting the number of attachments per cell reduces the size of a group, it effectively delays when groups begin to run out of space. A tree without degree caps can only divide 11 times before exceeding the available volume contained by a shell. In contrast, a tree with a degree cap of 3 can divide 19 times. On the 20th division it exceeds shells 14 and 15 after reaching $\approx 2*10^4$ cells, which is 5-10 times as many cells present in trees without degree caps when they reach volume constraints. A tree with degree cap of 4, however, only divides 14 times before encountering a limit at shells 8-11 and reaches about half as many cells ($\approx 10^4$). So the volume constraints are quickly encountered as degree caps increase above 3-- despite limiting the number of cellular attachments. Interestingly, degree caps of 2 produce filaments, a common biological shape, that are free of any volume constraints.
\par

\subsection*{Population simulations}
Thus far, we have examined the consequences on group reproduction of the death of a single cell. In practice, however, as groups reproduce and grow, mechanisms of cell death interact with geometric constraints to create a population of groups with a distribution of sizes. To determine this distribution, we simulate the population expansion from the first mutant capable of forming groups, {\it Node 0} using the numerical software {\tt MATLAB}(version 7.12.0.635  Natick, Massachusetts: The MathWorks Inc., 2011). All cells reproduce at the same rate, but cells cannot do so if they have reached the maximum degree (degree cap) or if the shell their offspring would belong to is full (volume constraints). Cell death occurs randomly and all cells are susceptible-- though we relax this assumption later to make younger cells less likely to die. If a cell dies then one of its links is randomly severed, biased by the number of cells along that branch (weighting). We simulate the population growth for 21 rounds of cell division and consider trees with different degree caps. 
\par
\begin{figure}[htbp]
\begin{center}
\includegraphics[scale=.55]{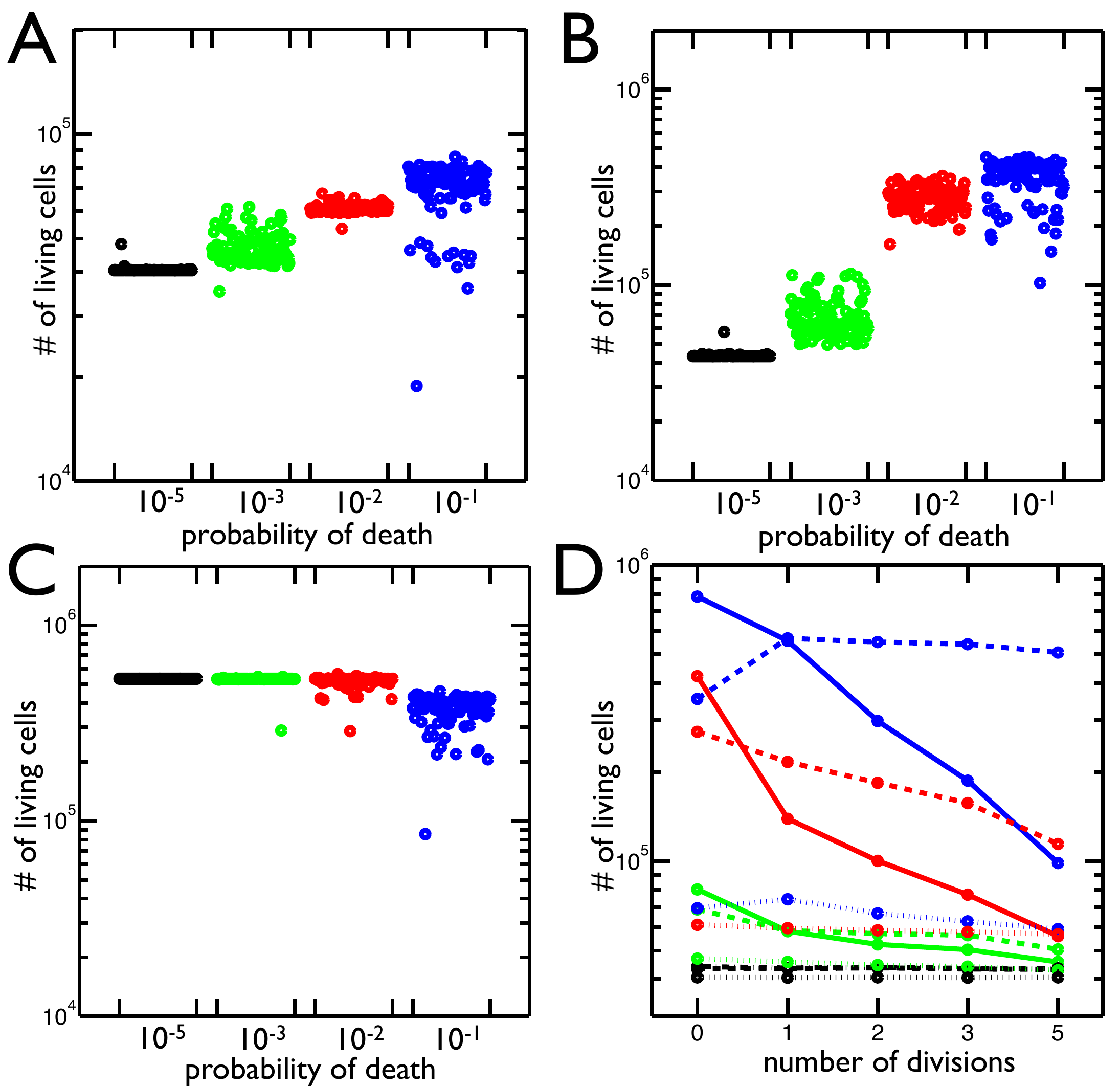}
\caption{{\bf Total number of cells resulting from different rates of cell death and geometric constraints.} {\bf A)} Total number of living cells in groups with maximum node degree 3 in 100 simulations at 4 different probabilities of cell death. The highest probability of death ($10^{-1}$ in blue) has the largest number of cells and the greatest variance in final cell number. {\bf B)} Same as A but with maximum node degree 4. Once again the highest probability of death ($10^{-1}$ in blue) produces the most cells. {\bf C)} The same as B but the volume constraints are removed, i.e. the maximum node degree is still 4 but there is no limit to the number of cells in each shell. The $10^{-1}$ probability of death no longer increases the population of cells. {\bf D)} The mean number of living cells when a cell's susceptibility to death is delayed by 0-5 rounds of division ($\sim$ time units) since its last division. The colors correspond to probabilities of death: $10^{-1}$ (blue), $10^{-2}$ (red), $10^{-3}$ (green), $10^{-5}$(black); and the line style represents the degree cap: no cap (solid), 4 (dashed), 3 (dotted). In trees with degree caps of 3 and 4, the highest probability of death results in even more cells when death is delayed one cell division but less as death is delayed further. In all other cases, delaying death results in less cells.}
\label{NumCells}
\end{center}
\end{figure}

The population simulations show that the total number of living cells increases with the probability of cell death (Figure \ref{NumCells}A \& B). This paradoxical result is a consequence of the constraints on cell reproduction due to degree caps and limited volume. For a degree cap of 3, cells that reach the maximum degree (3 in this case) stop reproducing. After 21 generations, many cells have reached the maximum degree and no longer contribute to the growth of the population. By dying, a link connecting two non-dividing cells is broken. This allows one cell to divide again and start a new branch that increases the population by more cells than the cost of the dead cell. Since a group with degree cap of 3 does not encounter volume limitations until the 20th cell division, near the end of the simulation, the volume constraint does not play a significant role in the increased cell population. In fact, it can be removed and the total number of cells still increases with higher rates of cell death. This is not true with groups that have a degree cap of 4 (Figure \ref{NumCells}C). The higher degree cap reduces the extent to which fixing a maximum number of attachments constrains the population while at the same time increases the strength of the volume constraints-- cells experience volume limitations by the 14th cell division. So, both degree cap and volume constraints allow groups to increase the number of living cells by increasing the frequency of cell death.
\par
In biological systems cell death may not be completely random but rather biased by age. Analytically, we showed that the age of the dead cell affects the expected sizes of group offspring. Here, we include a bias in the age at which cells die in the simulation by protecting dividing cells from death; cells cannot die until a set amount of time has past since their last cell division. We expect this to act similarly to decreasing the death rate because fewer cells are susceptible to death. As such, we predict that the longer death is delayed the lower the final population. Instead, we find that delaying death has a variety of effects depending on the degree cap and the frequency of cell death (Figure \ref{NumCells}D). The results match our expectations when the probability of death is low ($\le 10^{-2}$) or groups are not degree capped. In contrast, when the probability of death is high ($10^{-1}$), the number of cells in degree capped groups increases if death is delayed. This effect is strongest when death is delayed only one round of division, i.e. cells are susceptible when it has been at least one generation since their last cell division. As the delay gets longer the total number of cells decreases. For a degree cap of 4, delaying death for 5 rounds of division still produces more cells than when there is no delay, but this is not true for a degree cap of 3. Thus, delaying death has different effects depending on the probability of death, the length of the delay, and the maximum node degree. 
\par
\begin{figure}[htbp]
\begin{center}
\includegraphics[scale=.55]{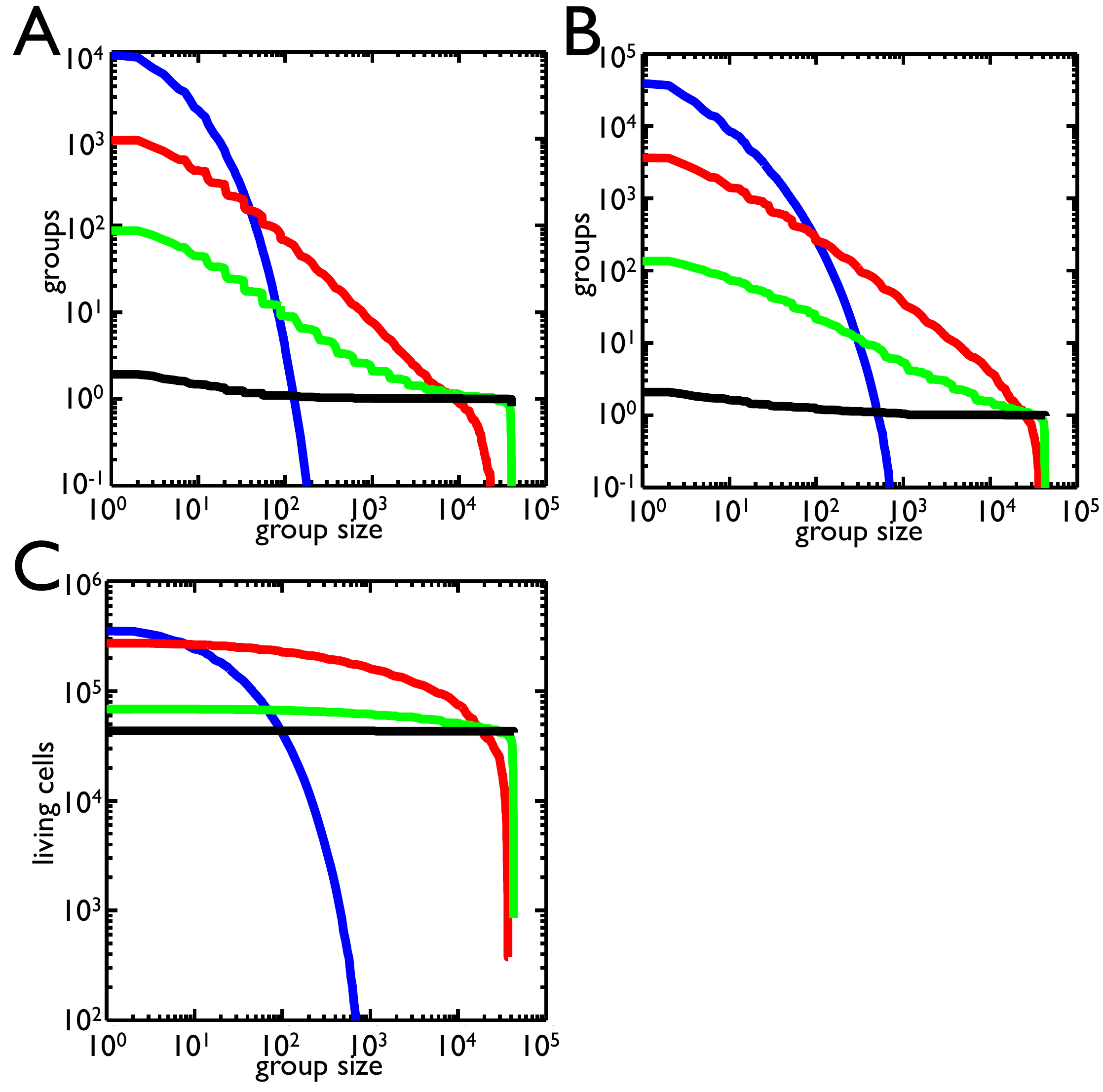}
\caption{{\bf Selection for group size.} {\bf A)} The number of groups that satisfy size thresholds are shown for a degree cap of 3 for different probabilities of death: $10^{-1}$ (blue), $10^{-2}$ (red), $10^{-3}$ (green), $10^{-5}$ (black). As the group size increases, the number of groups above threshold drops. Small group size favors higher probabilities of death while large group size favors low probability of death.  {\bf B)} Same as A but with a degree cap of 4. The range in which $10^{-1}$ is dominant has expanded and $10^{-2}$ does better at group sizes above $10^4$. {\bf C)} The number of cells within groups that satisfy size thresholds for a degree cap of 4 is shown for different probabilities of death (same color scheme). In contrast to B, the $10^{-2}$ probability of death has a much larger range in which it is best. Comparing B and C, there is a region between 10 and 100 cells in which the $10^{-1}$ probability of death produces more groups but fewer cells in those groups than $10^{-2}$.  }
\label{GrpSel}
\end{center}
\end{figure}

Due to volume constraints and degree caps, apoptosis can increase both the number of cells and the number of groups. Yet, the experimental regime rewarded cells that were in groups above a certain size-- this success might be measured as either the number of groups or the number of cells in groups. The frequency of cell death affects both the number and size distribution of groups. To find which apoptosis rate yields the most groups over different size thresholds, 
 we compute the average number of groups above threshold for different probabilities of death (Figure \ref{GrpSel}A for degree cap of 3 and \ref{GrpSel}B for degree cap of 4). For small group thresholds ($<$ 25 cells), the highest probability of death $10^{-1}$ produces the most group offspring. As the group threshold increases to $10^2$ cells, the probability of death that leaves the most group offspring decreases to $10^{-2}$. Larger size thresholds ($\ge 10^5$) effectively reward groups that never divide, and so the best strategy is to have the lowest probability of death (here, $10^{-5}$). These trends also hold if the degree cap is 4, but the higher probabilities of death ($10^{-1}$ and $10^{-2}$) dominate for greater ranges of size thresholds. Moreover, these trends are the same if fitness is determined not by the number of groups above threshold but rather the number of cells in those groups. Once again, the higher probabilities of death are successful for size thresholds from 1 to $\approx 10^4$. One notable difference is that for size thresholds between $10^1$ and $10^2$, the highest probability of death, $10^{-1}$, produces the most groups but not the most cells-- the $10^{-2}$ probability of death produces more cells in groups above threshold.
\par
\begin{figure}[htbp]
\begin{center}
\includegraphics[scale=.55]{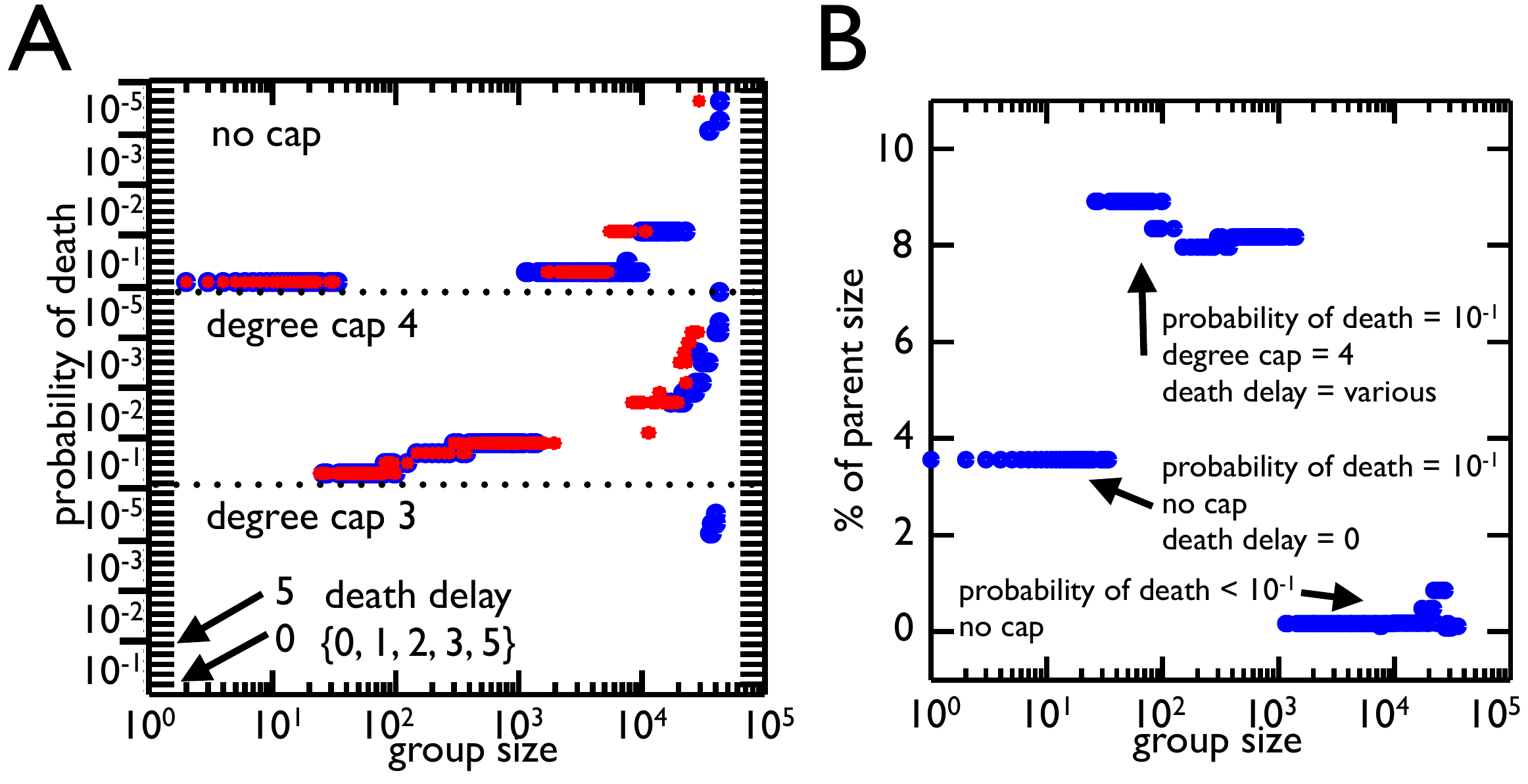}
\caption{{\bf Best strategies for different group size selection.}  {\bf A)} The strategy that yields the most group offspring is shown for each threshold of group size for different rounds of cell division (blue for 21, red for 19). The degree caps follow the same organization broken down by probability of death with ticks indicating death delay. In general the $10^{-1}$ probability of death for a degree cap of 4 and no cap is the best strategy for most group sizes. Once the group size gets large ($>10^4$) lower probabilities of death begin to win as it is advantageous not to divide large groups. {\bf B)} The average size of group offspring as a percent of parent size is shown for each optimal strategy from A. The values are all under $9\%$ and are much smaller than those experimentally observed. There was, however, only one strategy which left more symmetrical groups (a degree cap of 3 with the highest probability of death).}
\label{Best}
\end{center}
\end{figure}

In determining which apoptosis rate produces the most groups, we assumed that the probability of death is an evolvable trait. The same may be true of other features related to group organization or cell death such as degree cap and the age bias of cell death (the death delay). To find which combination of these traits, ``strategies", yields the most groups above threshold, we compare combinations of degree cap, probability of death, and death delay (Figure \ref{Best}A). For weak thresholds that permit small groups of less than 25 cells, the most groups are left by those without degree caps who have a probability of death of $10^{-1}$ and no death delay. This strategy also produced the most living cells without considering group size thresholds (Figure \ref{NumCells}D). For intermediate group thresholds between 25 and 1000 cells, a degree cap of 4 with a probability of death of $10^{-1}$ is best. As the size threshold increases within this range so does the optimal death delay.  For group size selection between $10^3$ and $10^4$ the best strategy shifts back to groups without degree caps who have a probability of death of $10^{-1}$ and death delays above 0. The largest group size selection ($>10^4$) finds the lowest probabilities of death with all degree caps doing well. In general, these results hold if the number of cell divisions in the simulations is reduced from 21 to 19.
\par
Finally, we consider how the best combinations of traits for different size thresholds fare in group offspring symmetry. We compute the average size of offspring group for the best strategies (Figure \ref{Best}B) and find that they produce much smaller offspring than the $30-40\%$ observed experimentally: $\approx 3\%$ of the parent's size for small group selection, $\approx 9\%$ for intermediate groups, and $< 1\%$ for large groups. Although they fall short, only a degree cap of 3 with the highest probability of death left more symmetrical group offspring ($\approx 20\%$). The symmetry of offspring did not compensate for the limits such a stringent degree cap places on population size.

\section*{Discussion}
An experiment exploring the emergence of multicellularity observed the rapid evolution of groups from unicellular precursors in the yeast {\it Saccharomyces cerevisiae} when cultures were placed under selection for rapid settling through liquid medium \cite{Ratcliff2012}. Soon after the establishment of groups, cells also evolved a higher rate of apoptosis. Elevated cell death clearly lowers cell viability, but it would also seem to lower {\it group} viability. This is because settling selection favors large clusters and cell death facilitates group division, and thus size reduction. Why would natural selection favor elevated-- as opposed to reduced-- levels of apoptosis?  Here we show that the organization of the group and the constraints imposed by its geometry are instrumental in understanding the functional consequences  of apoptosis. By increasing the frequency of cell death, both the number of cells and groups can increase. Thus, a trait which is harmful to the cells that express it (they die) acts as a form of suicidal altruism and is beneficial to both the long-term number of cells and group entities once the group structure is considered. Furthermore, this trait may play a key role in the evolutionary transition to multicellularity.
\par
With the transition from unicellularity to multicellularity there is an important shift in the level of organization and individuality \cite{Smith1997,Calcott2011,Okasha2005}.  A key requirement for multicellularity is formation of a cohesive group of cells. Group formation offers distinct advantages over a strictly solitary lifestyle such as protection from predation \cite{Keesin1996,Boraas1998}, access to new niches \cite{Bonner1998}, and survival in harsh environments \cite{Smukalla2008}. However, for groups to qualify as units of selection, they must also possess the capacity to beget group offspring \cite{Okasha2009,Rainey2010,Libby2013}. In this experimental yeast system, clusters grew in size and as a result of cell death or physical strain they fragmented and thereby reproduced. Thus, the yeast simultaneously evolved group formation and a mode of reproduction {\it de novo}. The later evolution of increased cell death led to more frequent cluster reproduction, thereby, linking reproductive self-sacrifice at the lower-level to fecundity at the higher-level. From a certain perspective, the fitness of the apoptotic lower-level units is subjugated to elevate the fitness of higher-level units, which is taken to be a hallmark of an evolutionary transition in individuality \cite{Smith1997,Calcott2011,Okasha2009,Michod1999,Michod2003}. 
\par
Interestingly evolution of increased cell death also acts to stabilize the transition to multicellularity. If a cell with a higher rate of death were to leave the context of its collective, it would not fare well in competition with other cells who never formed groups (and never evolved greater apoptosis).  In this way, the trait ratchets cells into a multicellular lifestyle by making them less competitive with their unicellular ancestors. This prevents abandonment of the collective and reversion to unicellularity. By tying the fate of cells to the fate of their groups, such context dependent traits stabilize primitive multicellular forms.
\par
The amount of stabilization provided by a context dependent trait would likely depend on the fitness tradeoff between unicellular and multicellular life. More stabilization is expected from traits that severely hamper the fitness of cells outside the group context. It is unknown whether such stabilizing traits are common but with the yeast system analyzed in this paper there is robust selection for increased apoptosis rates. Rather than finding a narrow range of conditions that selected for moderately higher rates of cell death, we found strong selection for high rates of cells death ($1-10\%$) across a wide spectrum of group size thresholds. In fact, the only regime where increased cell death does not succeed is when groups need to be close to the maximum possible size. This regime selects for the lowest cell death rates and results in a single group encompassing the entire population. Otherwise, when size selection required minimum group sizes from 0 to $10^4$, high rates of cell death allowed cells to circumvent limitations imposed by geometry. Interestingly, these limitations were of two different classes: limits to the number of connections when the maximum degree is 3 and limits to space when the maximum degree is 4 and higher. As a consequence, a gamut of different tree geometries encounter limitations to growth that robustly select for high rates of apoptosis.
\par 
Our model implicitly assumes that the environment in which cells and groups grow is nutrient rich, and that the death of a cell provides the possibility for replacement by future cells. This allows apoptosis to overcome the cost of sacrificing a cell through the benefit of additional cellular reproduction. If, instead, the environment were nutrient poor and death of a cell did not guarantee replacement, then high rates of apoptosis would encounter an additional cost not reflected in our model, and would likely be less successful. It is possible that the model could be modified to consider cell survival as a function of crowdedness rather than cell fecundity. Cells in more crowded areas have less access to nutrients and by dying could create more access for neighbors, potentially improving their survival. These considerations lie outside the scope of this paper. In the experimental regime, as in the model, populations were grown in nutrient rich environments and so increased apoptosis led to both higher group and cell number. Still, it is important to recognize that the fitness consequence of traits depend on both the environment established by the group as well as its external environment.
\par
Considering the fitness implications of group geometry reveals that the group represents a novel, dynamic environment, one constructed bottom-up by individual cells. As such, variations in cellular physiology affect the geometry of the cluster, which in turn affects cell growth and survival. For example, if a cell has a morphology that only permits three connections to other cells, then the maximum possible cluster size will be much smaller than a cellular morphology that permits four connections. Similarly, different group formations impose different selective pressures on the cells within groups. The difference between three and four connections determines when cells will run out of space within the group. Although we considered a simple model with identical cells defined by just a few properties (maximum degree, apoptosis rate, and death delay), we found that these traits interact in surprising ways. For instance, increasing cell death increased the number of living cells but delaying death for cells-- effectively reducing the apoptosis rate-- had contrasting effects depending on the maximum degree and duration of the delay.  
\par
We only investigated how altering cell death affects cluster size and the number of cells/clusters in the population, but it is possible that cells could evolve different shapes, sizes, or behaviors which modify whole group-level traits. In fact, recent work has shown that strong selection for faster settling results in the evolution of larger, more elongate cells, which increase both group size and settling speed  \cite{Ratcliff2013}.  The environment faced by cells in clusters is not uniform: cells in the interior should experience a lower concentration of resources (as they must diffuse past other yeast that are consuming them) and higher concentrations of waste products. These environmental gradients provide robust cues that could allow a cell to determine its position within the cluster. As cells change their location within the geometry and experience different internal environments, it may be advantageous to adopt different strategies or forms. This raises the possibility that selection can favor location-specific morphological or behavioral differentiation.   Indeed, this may provide an evolutionary origin to primitive multicellular developmental programs.
\par

\newpage


\newpage
\section*{Supplement: Dynamic Programming Results}

In our model, clusters grow over $T$ time steps and gravity selection then occurs. At the beginning of each time step, cluster division can occur. Following division, clusters grow, such that any cluster of size $x$ cells will be $G(x)$ cells by the end of the time step. Time point $t=0$ marks the beginning of the first time step and time $t=T$ marks the point of gravity selection.  Upon selection, a cluster of $x$ cells survives with probability $S(x)$, which we take to be a non-decreasing function. The maximal reproductive output for a cluster of size $x$ at time $t$ is given by $F(x,t)$. For us, this output function is simply a means to determine the optimal way for clusters to split, which can depend on both size and time. In our scheme, a cluster of $x$ cells can split into two clusters of sizes $p$ and $x-p$ (where $0 \leq p \leq x/2$). Because it is possible for the cluster not to split (i.e., if $p=0$), we can simultaneously address the optimal rate of division along with optimal (a)symmetry.
\par
A backwards recursion for maximal reproductive output can be formulated:
\begin{equation}
\label{dprecursion2}
F(x,t)=\max_{0 \leq p \leq {x/2}}(F(G(p),t+1)+F(G(x-p),t+1))
\end{equation}
\par
Suppose that $G(x)=ax$ (where $a$ is an integer greater than unity) and $F(x,t+1)$ is concave.  We note that $F(x,t+1)$ is only defined for integer values of $x$, so the standard requirement ($\frac {d^2F(x,t+1)}{dx^2}\leq0$) is replaced by the following condition:
\begin{equation}
\label{concavity}
(F(x+1,t+1)-F(x,t+1))-(F(x,t+1)-F(x-1,t+1)) \leq 0
\end{equation}
If condition \ref{concavity} holds for all values of $x$, then we have the following proposition.
\vspace{.5cm}
\begin{mdframed}
{\bf Proposition}
\par
Given that $F(x,t+1)$ is concave by condition \ref{concavity}; for all integer values of $p$, where $0 \leq p \leq \frac{x}{2}$:
\begin{equation}
\label{GeneralJensen}
F(ap,t+1)+F(a(x-p),t+1) \leq \left\{ 
  \begin{array}{l l}
    2F\left(\frac {ax} {2},t+1\right) & \quad \text{if $x$ is even}\\
    F\left(\frac {a(x+1)} {2},t+1\right) + F\left(\frac {a(x-1)} {2},t+1\right) & \quad \text{if $x$ is odd}
  \end{array} \right.
\end{equation}
\end{mdframed}

\vspace{.5cm}

\begin{mdframed}
{\bf Proof}
\par
Here we use a proof by induction. Consider the case where $x$ is even. Let $n=a\left(\frac{x}{2}-p\right)$ (for any defined value of $p$, $n$ will be some non-negative integer value). Condition \ref{GeneralJensen} can be rewritten as:
\begin{equation}
\label{GeneralJensen2}
F\left(\frac {ax} {2},t+1\right) \geq \frac {F\left(\frac {ax} {2} - n,t+1\right) + F\left(\frac {ax} {2} + n,t+1\right)}{2}
\end{equation}
Here, we will consider all non-negative integer values of $n$ (even those that don't correspond to an integer value of $p$). Condition \ref{GeneralJensen2} clearly holds for $n=0$.  Additionally, it holds for $n=1$ because condition \ref{concavity} can be rewritten (with $x$ replaced by $\frac{ax}{2}$) as
\begin{equation}
\label{GeneralJensen3}
F\left(\frac {ax} {2},t+1\right) \geq \frac {F\left(\frac {ax} {2} - 1,t+1\right) + F\left(\frac {ax} {2} + 1,t+1\right)}{2}
\end{equation}
We now assume that condition \ref{GeneralJensen2} holds for $n$ and show that it must hold for $n+1$. If $F(x,t+1)$ is a concave function, then we are guaranteed
\begin{equation}
\label{GeneralJensen4}
F(x-1,t+1) \leq 2F(x,t+1) - F(x+1,t+1)
\end{equation}
\begin{equation}
\label{GeneralJensen5}
F(x+1,t+1) \leq 2F(x,t+1) - F(x-1,t+1)
\end{equation}
Using conditions \ref{GeneralJensen4} and \ref{GeneralJensen5}, the following holds:
\begin{equation}
\label{GeneralJensen6}
\raggedleft {
\begin{split}
\frac {F\left(\frac {ax} {2}+(n+1),t+1\right)+F\left(\frac {ax} {2}-(n+1),t+1\right)}{2} & \leq F\left(\frac {ax} {2}+n,t+1\right)+F\left(\frac {ax} {2}-n,t+1\right) \\
& - \frac {F\left(\frac {ax} {2}+(n-1),t+1\right)+F\left(\frac {ax} {2}-(n-1),t+1\right)}{2}
\end{split}
}
\end{equation}
The following condition holds
\begin{equation}
\label{GeneralJensen7}
\frac {F\left(\frac {ax} {2}+n,t+1\right)+F\left(\frac {ax} {2}-n,t+1\right)}{2}- \frac {F\left(\frac {ax} {2}+(n-1),t+1\right)+F\left(\frac {ax} {2}-(n-1),t+1\right)}{2} \leq 0
\end{equation}
To show condition \ref{GeneralJensen7}, we note that condition \ref{GeneralJensen5} ensures
\begin{equation}
\label{GeneralJensen8}
\raggedleft {
\begin{split}
F\left(\frac {ax} {2}+n,t+1\right)-F\left(\frac {ax} {2}+(n-1),t+1\right)+F\left(\frac {ax} {2}-n,t+1\right)-F\left(\frac {ax} {2}-(n-1),t+1\right) \leq \\
F\left(\frac {ax} {2}+(n-1),t+1\right)-F\left(\frac {ax} {2}+(n-2),t+1\right)+F\left(\frac {ax} {2}-n,t+1\right)-F\left(\frac {ax} {2}-(n-1),t+1\right)
\end{split}
}
\end{equation}
However condition \ref{GeneralJensen5} also ensures
\begin{equation}
\label{GeneralJensen9}
\raggedleft {
\begin{split}
F\left(\frac {ax} {2}+(n-1),t+1\right)-F\left(\frac {ax} {2}+(n-2),t+1\right)+F\left(\frac {ax} {2}-n,t+1\right)-F\left(\frac {ax} {2}-(n-1),t+1\right) \leq \\
F\left(\frac {ax} {2}+(n-2),t+1\right)-F\left(\frac {ax} {2}+(n-3),t+1\right)+F\left(\frac {ax} {2}-n,t+1\right)-F\left(\frac {ax} {2}-(n-1),t+1\right)
\end{split}
}
\end{equation}
And the same substitution can be repeatedly applied, which yields
\begin{equation}
\label{GeneralJensen10}
\raggedleft {
\begin{split}
F\left(\frac {ax} {2}+n,t+1\right)-F\left(\frac {ax} {2}+(n-1),t+1\right)+F\left(\frac {ax} {2}-n,t+1\right)-F\left(\frac {ax} {2}-(n-1),t+1\right) \leq \\
F\left(\frac {ax} {2}-(n-1),t+1\right)-F\left(\frac {ax} {2}-n,t+1\right)+F\left(\frac {ax} {2}-n,t+1\right)-F\left(\frac {ax} {2}-(n-1),t+1\right) = 0
\end{split}
}
\end{equation}
Thus, condition \ref{GeneralJensen7} follows.
\par
Condition \ref{GeneralJensen7} shows that condition \ref{GeneralJensen6} can be rewritten as
\begin{equation}
\frac {F\left(\frac {ax} {2}+(n+1),t+1\right)+F\left(\frac {ax} {2}-(n+1),t+1\right)}{2} \leq \frac {F\left(\frac {ax} {2}+n,t+1\right)+F\left(\frac {ax} {2}-n,t+1\right)}{2}
\end{equation}
Given that we are assuming that condition \ref{GeneralJensen2} holds for $n$, it now follows
\begin{equation}
\frac {F\left(\frac {ax} {2}+(n+1),t+1\right)+F\left(\frac {ax} {2}-(n+1),t+1\right)}{2} \leq F\left(\frac {ax} {2},t+1\right)
\end{equation}
Thus, condition \ref{GeneralJensen2} holds for $n+1$.  Thus, this condition will hold for all integer values of $n$, which certainly ensures that it will hold for all integer values of $p$ (where $0 \leq p \leq \frac{x}{2}$). The case where $x$ is odd follows a similar argument. This completes the proof.
$\blacksquare$
\end{mdframed}
\vspace{.5cm}
\par
Condition \ref{GeneralJensen2} is essentially an instance of Jensen's inequality.  Using Eq. \ref{dprecursion2} and condition \ref{GeneralJensen},
\begin{equation}
\label{Jensen2}
F(x,t) = \left\{ 
  \begin{array}{l l}
    2F\left(\frac {ax} {2},t+1\right) & \quad \text{if $x$ is even}\\
    F\left(\frac {a(x+1)} {2},t+1\right) + F\left(\frac {a(x-1)} {2},t+1\right) & \quad \text{if $x$ is odd}
  \end{array} \right.
\end{equation}
Suppose that $x$ is even; then Eq. \ref{Jensen2} ensures
\begin{equation}
\label{Jensen3}
\raggedleft {
\begin{split}
(F(x,t)-F(x-1,t))-(F(x-1,t)-F(x-2,t)) & = 2F\left(\frac {ax} {2},t+1\right) - 2F\left(\frac {ax} {2},t+1\right) \\
& - 2F\left(\frac {a(x-2)} {2},t+1\right) + 2F\left(\frac {a(x-2)} {2},t+1\right) \\
& = 0
\end{split}
}
\end{equation}
Suppose that $x$ is odd; then Eq. \ref{Jensen2} ensures
\begin{equation}
\label{Jensen4}
\raggedleft {
\begin{split}
(F(x,t)-F(x-1,t))-(F(x-1,t)-F(x-2,t)) & = F\left(\frac {a(x+1)} {2},t+1\right) + F\left(\frac {a(x-1)} {2},t+1\right) \\
& - 2F\left(\frac {a(x-1)} {2},t+1\right) - 2F\left(\frac {a(x-1)} {2},t+1\right) \\
& + F\left(\frac {a(x-1)} {2},t+1\right) + F\left(\frac {a(x-3)} {2},t+1\right)\\
& = \left[F\left(\frac {a(x+1)} {2},t+1\right) - F\left(\frac {a(x-1)} {2},t+1\right)\right] \\
& - \left[F\left(\frac {a(x-1)} {2},t+1\right) - F\left(\frac {a(x-3)} {2},t+1\right) \right]
\end{split}
}
\end{equation}
Because $F(x,t+1)$ is concave by assumption, this means that the right-hand side of Eq. \ref{Jensen4} is less than or equal to zero; thus, for all relevant $x$
\begin{equation}
\label{Jensen5}
(F(x,t)-F(x-1,t))-(F(x-1,t)-F(x-2,t)) \leq 0
\end{equation}
\par
This means that if $F(x,t+1)$ is concave over integer values of $x$, then $F(x,t)$ will be as well.  Thus, if $F(x,T)=S(x)$ is concave, then $F(x,t)$ will be concave for all values of $t$.  This means that there will be no better strategy than splitting the group into two equal pieces (or as close as possible).
\par
If it were possible for $F(x,t+1)$ to be convex for all integer values ($(F(x+1,t+1)-F(x,t+1))-(F(x,t+1)-F(x-1,t+1)) \geq 0$), then a similar argument would show that $F(x,t)$ will be convex as well.  In such a case, it would be best for the cluster not to split at all ($p=0$ is the optimal value). Given that $S(x)$ is a probability (and thus bounded at unity) this function cannot be convex for all values of $x$. Therefore, if $F(x,T)=S(x)$, then it will generally not be the case that $F(x,t)$ is convex for arbitrary values of $x$ and $t$. It is possible for $S(x)$ (and $F(x,t)$) to be convex for some values of $x$ and concave for other values. In such a case, it is possible for the optimal division strategy to depend on size and time. For instance, in Figure 7, the results of a program to calculate the optimal division values ($p$ as a function of $x$ and $t$) using Eq. \ref{dprecursion2} are given for a set of different functions for $F(x,T)$ (in the figure, we do impose a maximum size for a cluster). Thus, we see that there are circumstances where a cluster should not divide at certain sizes and divide evenly at other sizes.

\newpage
\section*{Figures}


\begin{figure}[htbp]
\begin{center}
\includegraphics[width=.9 \textwidth]{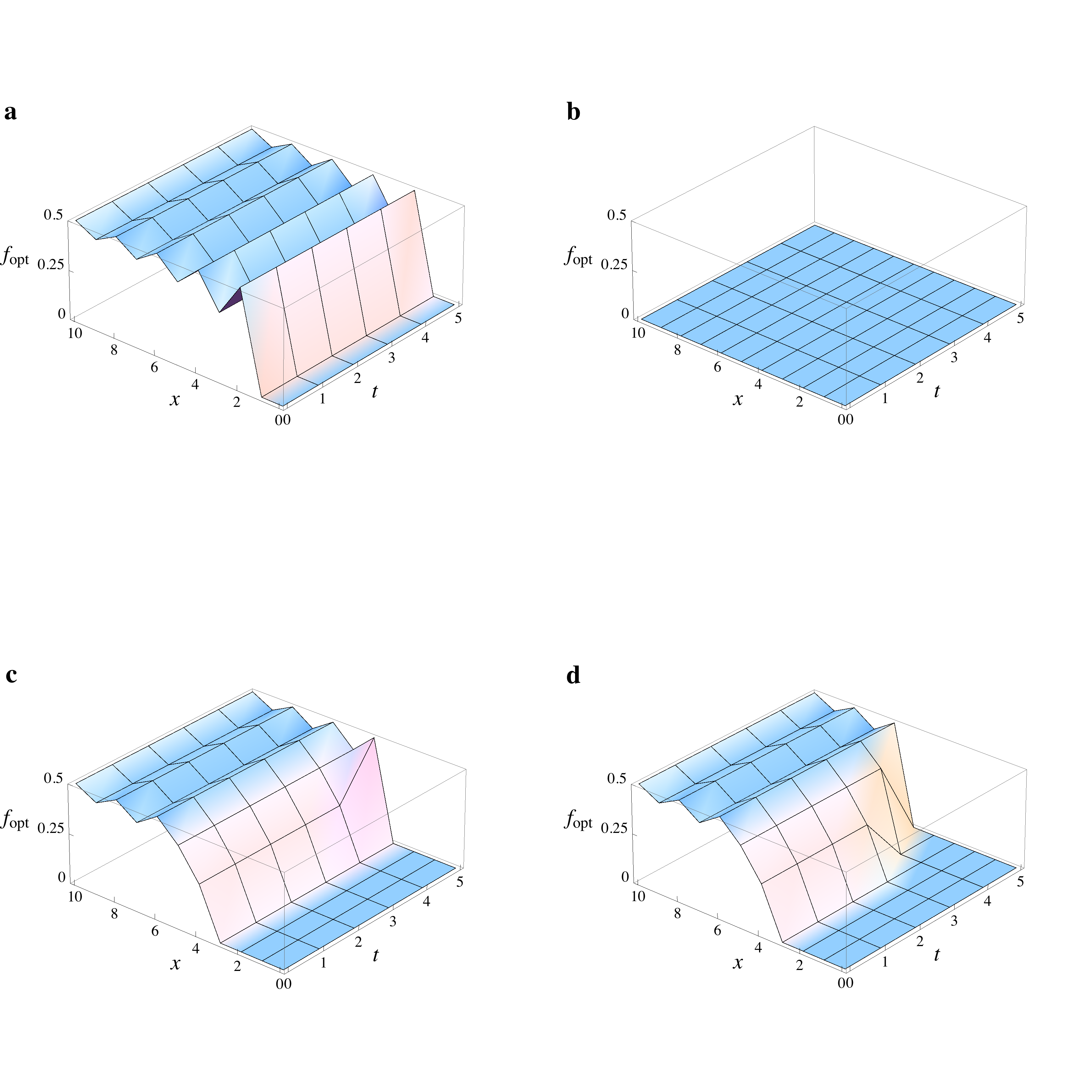}
\caption{{\bf Optimal propagule sizes using the dynamic programming approach}. Here we programmed the recursion from Eq. \ref{dprecursion2} and solved for the optimal division $p_{opt}$ as a function of cluster size ($x$) and time ($t$). In our graphs we focus on a small range of sizes ($0 \leq x \leq 10$) over five time steps. For all runs we assume $G(x)=2x$ (clusters double every time step). In the program we set a maximum cluster size (of $x_{max}=350$; note, the largest a cluster could get within our focal range would be $(10)(2^5)=320$). We vary the function $F(x,T)$ in these graphs and plot the optimal division size as the smaller fraction of a cluster after the split ($f_{opt}$). Note for $x=0$, we set $f_{opt}=0$ (although actually the optimal fraction is undefined) and for $x=1$, $f_{opt}$ must be 0. Whenever distinct fractions give equivalent optimal strategies, the smallest fraction is plotted. {\bf a)} Here we have a purely concave function $F(x,T)=S(x)=(1-e^{-0.1x})$, and we see that the optimal strategy is to split the cluster into two equal pieces. Of course, for clusters with an odd number of cells, this is impossible, but the optimal strategy is to divide the cluster as evenly as possible (e.g., a cluster of size 5 gets split into a cluster of size 3 and one of size 2, for $f_{opt}=\frac {2} {5} = 0.4$). {\bf b)} Suppose the survival function is purely concave ($S(x)=(1-e^{-0.1x})$ as before), but now maximal reproductive output is measured in terms of number of {\it cells} surviving selection, and not the number of clusters. Here, $F(x,T)=xS(x)=x(1-e^{-0.1x})$. In this case, it is optimal to avoid splitting under all conditions in our range. {\bf c)} In cases where the survival function flips concavity across our range, optimal division can depend on size and time. Here $F(x,T)=S(x)=\frac{x^h}{y^h+x^h}$, where $y=5$ and $h=2$. {\bf d)} The same function is used here as in part (c), but $h=10$.}
\label{SupFig}
\end{center}
\end{figure}

\end{document}